\documentclass[runningheads]{llncs}
\usepackage[T1]{fontenc}
\usepackage{newtxtext,newtxmath}
\DeclareMathAlphabet{\mathcal}{OMS}{cmsy}{m}{n}
\DeclareSymbolFont{letters}{OML}{txmi}{m}{it} 

\usepackage{graphicx}
\usepackage{tikz}
\usetikzlibrary{shapes.geometric, arrows}
\usepackage{enumerate}
 \usepackage{enumitem}
\usepackage{amsmath}
\usepackage{multirow}
\begin{document}
\title{Towards an Enforceable GDPR Specification}
%
%
\author{François Hublet\and
Alexander Kvamme\and
Sr\dj{}an Krstić}
\authorrunning{F. Hublet et al.}
%
\institute{Departement of Computer Science, ETH Zürich
\email{\{francois.hublet,srdan.krstic\}@inf.ethz.ch}}
\maketitle              
\begin{abstract}
While Privacy by Design (PbD) is prescribed by modern privacy
regulations such as the EU's GDPR, achieving PbD in real
software systems is a notoriously difficult task. One emerging
technique to realize PbD is \emph{Runtime enforcement} (RE), in which an \emph{enforcer}, loaded with a specification of a system's privacy requirements, observes the actions performed by the system and instructs it to perform actions that will ensure compliance with these requirements at all times.

To be able to use RE techniques for PbD,
privacy regulations first need to be translated into an enforceable specification. In this paper, we report on our ongoing work in 
formalizing the GDPR. We first present a set of 
requirements and an iterative methodology for creating 
enforceable formal specifications of legal provisions. Then, we 
report on a preliminary case study in which we used our 
methodology to derive an enforceable specification of part of 
the GDPR. Our case study suggests that our methodology can be effectively used to develop accurate enforceable specifications.
\end{abstract}

\section{Introduction}

Ensuring compliance of software systems with privacy regulations 
is a notoriously challenging task. With mounting evidence that 
current regulations such as the EU's General Data Protection 
Regulation (GPDR) are poorly implemented by a majority of 
data controllers~\cite{Bollinger2022}, the capacity of ex-post enforcement and fines 
to bring about widespread compliance 
appears limited. This calls for 
a methodology that relies
on principled approaches to \emph{design} and \emph{certify} software systems 
to adhere to regulations, in the spirit of Privacy by Design (PbD)~\cite{Cavoukian2009}.

\emph{Runtime enforcement}%
\footnote{Note that the term `enforcement' has a different meaning in 
legal and computer science contexts. In the former, it typically refers to
the actions taken by state officials to end or penalize a past or ongoing
violation of the law. In the latter, 
`(runtime) enforcement' refers to a process of ensuring compliance 
with a policy at any time, \emph{preventing} policy 
violations rather than compensating for them. 
In the following, `enforcement' shall be understood in this latter sense.}
(RE) is one such approach. In RE, a formal specification is input to a software system, called an \emph{enforcer}, that observes the actions performed by a System under Scrutiny (SuS). In addition to observing the SuS's actions, the enforcer can also send commands to the SuS, typically instructing the SuS to prevent or cause certain actions. By observing the SuS's actions and responding with appropriate commands, the enforcer seeks to ensure that the behavior of the SuS adheres to its specification at all times. A specification for which such an enforcer exists is called \emph{enforceable}.

In the context of privacy, typically regulations (e.g., the GDPR), 
must be enforced. 
Therefore, to use RE for enforcing privacy in software systems one must
\emph{formalize} privacy regulations into \emph{enforceable specifications}. 
This entails \emph{understanding} the regulations, making them \emph{precise}
in the context of specific software system actions, making them \emph{formal} (i.e., readable by an enforcer), and, finally, \emph{enforceable}. 
In this paper, we report on our ongoing work in 
formalizing GDPR that is both close to the letter of the law and enforceable.
Our work can serve as a common ground for both computer
scientists and legal experts to collaborate in achieving PbD.

This paper is organized as follows. 
First, we review existing work on formalizing privacy regulations, 
especially the GDPR (Section~\ref{sec:related work}) and
briefly introduce RE (Section~\ref{sec:enforcement}). 
We then present four requirements and an iterative methodology for 
obtaining an enforceable specification of regulations in general (Section~\ref{sec:methodology}). Our methodology is applicable both to regulations for which no specification exists and those for which 
an existing \emph{non-enforceable} (or inaccurate) specification exists.
Finally, we report on a case study in which we applied our methodology to 
convert a series of GDPR provisions formalized in DAPRECO~\cite{Robaldo2020}
into enforceable specifications (Section~\ref{sec:case study}). 
In the course of our case study, we discovered a number of inaccuracies in DAPRECO that can be, at least in part, linked to Robaldo et al.'s choice of a formalism and methodology. We discuss how our methodology can be used to alleviate the risk of such inaccuracies while additionally providing out-of-the-box support for RE in real systems. In conclusion, we reflect on the limitations of our approach and open questions (Section~\ref{sec:conclusion}).

\section{Related work} 
\label{sec:related work}

While some of the first formal specifications of legal provisions~\cite{Sergot1986} date back
to the 1980s and there is a related active field 
at the interplay of logic, linguistics, computer science, and law
(see, e.g.,~\cite{Gabbay2010}), formal specifications of legal provisions amenable
to formal reasoning are still relatively rare. Recently, tax codes were formalized using special-purpose programming
languages~\cite{Merigoux2021,Huttner2022}. The complexity of tax codes lies in
complex conditional structures depending on a large number of variables and 
conditions, requiring the use of special (typically, default~\cite{Antoniou1999}) logics. 
However, tax codes lack most of the temporal and system-specific
dimensions that are intrinsic to privacy regulations.

The runtime verification community~\cite{Bartocci2018}
has seen an increasing number of efforts to 
detect (or \emph{monitor})
violations of complex policies in large-scale systems.
These efforts have brought to the fore a number of challenges
in terms of both policy engineering and the design of the
interaction between the SuS and monitors~\cite{ColomboP18}. One
of the most comprehensive studies to date is Basin et al.'s monitoring
of several millions of log entries from the ``Internet Computer,'' a distributed Web3 platform,
against large temporal-logic policies with over 1,000 binary operators~\cite{BasinDKPRST23}. The latter study required very
careful policy engineering, as temporal logic policies had to be extracted from an informal description of protocol properties provided by engineers. To the best of our knowledge, no such large-scale study has been conducted yet with \emph{legal} provisions.

In the last two decades, smaller portions of privacy laws were formalized, often
based on ad-hoc interpretations tailored to specific application domains. In an
early work, Lam et al. formalized part of the US Health Insurance  Portability
and Accountability Act (HIPAA) using stratified Prolog~\cite{Lam2009}. 
Arfelt et al.~\cite{Arfelt2019} 
formalized data subject rights and monitored them in industrial logs. 
Hublet et al.~\cite{Hublet2023} showed how to enforce a similar core of GDPR 
provisions in web applications and extended Arfelt et al.'s study to support runtime enforcement~\cite{Hublet2024}. Palmirani and Governatori used LegalRuleML~\cite{Palmirani2011} and the PrOnto ontology~\cite{Palmirani2018} to model GDPR provisions~\cite{Palmirani2018b}. Bonatti et al. formalized selected GDPR-related constraints on business processes using the Web Ontology Language OWL2~\cite{Bonatti2020}. On another line of work, a large number of publications (see the survey~\cite{Esteves2022}) have focused on defining GDPR-compatible \emph{policy languages} that can be used to describe legal flows of information depending on, e.g., user consent, and help users define their own preferences. Finally, Torre et al.~\cite{Torre2021} have translated GDPR concepts into full-fledged ontologies, which, however, lack an associated specification of legal provisions.

Only a handful of previous works have attempted to formalize entire privacy regulations in a more literal way, capturing legal concepts into an appropriate ontology and converting individual paragraphs of the original legal document into logical formulae. In these respects, two series of works stand out. The first one is DeYoung et al.'s extensive formal specification of those parts of the Gramm-Leach-Bliley Act (GLBA) and HIPAA~\cite{DeYoung2010,DeYoung2010b} that defined ``operational requirements'' of systems. DeYoung et al. use an extension of Least Fixed Point logic (LFP) to encode these requirements into a well-documented set of LFP formulae. The second series of work is Robaldo, Bartolini et al.'s DAPRECO knowledge base~\cite{Robaldo2020,Robaldo2020b}, which provides the most comprehensive formal specification of the GDPR to date. The authors provide an ontology and a set of over 900 formulae that they claim to cover most of the GDPR except articles 51--76. A small portion of this specification has been validated through interdisciplinary collaboration with legal experts~\cite{Bartolini2018,Bartolini2019}. None of these two series of work consider enforcement.

\section{Runtime enforcement}
\label{sec:enforcement}

In this section, we briefly overview runtime enforcement (RE). We start from a system model that we assume a system under scrutiny (SuS) implements. This model is a simplified version of a model 
introduced in our previous work~\cite{Hublet2024}. We then present the specification language that our enforcer supports.

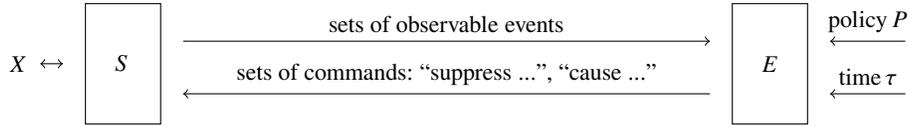
\begin{figure}[t]
  \begin{center}
    \begin{tikzpicture}
      \node at (-3.7,0.2) {$X$};
      \draw (-2.8,-0.6) rectangle (-1.8,1);
      \node at (-2.3,0.2) {$S$};
      \draw (5.8,-0.6) rectangle (6.8,1);
      \node at (6.3,0.2) {$E$};
      \draw[->] (-1.5,0.5) to node[above] {sets of observable events} (5.5,0.5);
      \draw[<-] (-1.5,-0.2) to node[above] {sets of commands: ``suppress ...'', ``cause ...''} (5.5,-0.2);
      \draw[<-] (7.1,0.5) to node[above] {policy\,$P$} (8.1,0.5);
      \draw[<->] (-3.4,0.2) to node[above] {} (-3.1,0.2);
      \draw[<-] (7.1,-0.2) to node[above] {time\,$\tau$} (8.1,-0.2);
    \end{tikzpicture}
  \end{center}
\vspace{-5ex}
  \caption{System model for runtime enforcement\label{fig:enforcement}}
\vspace{-3ex}
\end{figure}

\paragraph{System model.}
Figure~\ref{fig:enforcement} shows how an enforcer $E$ supervises a SuS $S$,
which interacts with an environment $X$ that $E$ cannot control.
$E$ must ensure that the sequence of actions executed by $S$
complies with a given policy $P$. 
An example of such a policy is
$$P_1=\text{``any use of a data item in a system has been preceded by user consent.''}$$
$E$ also has access to the current time $\tau$ from some reference clock.

During its execution, $S$ reports \emph{events} to $E$. 
An event 
is $e(a_1,\dots,a_k)$ where $e$ is called the `event name' and 
$a_1,\dots,a_k$ are `arguments.' An example of such an event is $$\mathsf{uses}
(``\mathtt{website.com}",``\mathtt{birthday}",``\mathtt{Alice}",``\mathtt{advertisement}"),$$ meaning 
``the application hosted at \texttt{website.com} is using user \texttt{Alice}'s birthday for advertisement purposes.''
In general, these events are taken from some appropriate ontology and 
encode specific actions of $S$. Events (and the corresponding actions) s
that $S$ reports to $E$ are called \emph{observable}. 
$E$ records events in a \emph{log} 
together with the current time. The log is a sequence $\sigma = ((\tau_1,D_1),\dots,(\tau_k,D_k),\dots)$,
where $\tau_1,\dots,\tau_k$ are timestamps 
and $D_1,\dots,D_k$ are sets of events. 
Each pair $(\tau_i,D_i)$ is called a \emph{time-point}.

$E$ can then emit \emph{commands}
instructing $S$ to cause and suppress some of its actions. An action
that can be caused is represented by a \emph{causable event}; 
an action that can be suppressed is represented by a \emph{suppressable} 
event. Which events are causable and suppressable depends on the system's
functionality and the way it implements its interface with the enforcer. 
For instance, if $S$ is processing data, \emph{data usage} can typically 
be made \emph{suppressable} by ensuring that $S$ always asks $E$ 
for permission \emph{before} processing a data item.
On the other hand, \emph{erasure of data} can typically be made \emph{causable} provided that $S$ provides the enforcer $E$ with an interface to execute the erasure of specific data items (e.g., from its database).
In our model~\cite{Hublet2024}, we allow $E$ to send commands to $S$ both 
in response to the logging of sets of events (``reactively'')
and on its own initiative (``proactively'').

\paragraph{Specification.} The specification needed to perform enforcement in the above model consists of three components:
\begin{enumerate}
    \item An ontology listing the available event names and the types of their arguments and describing their high-level meaning in terms of system actions.
    \item A mapping of every event name to a set of attributes among `observable', `causable,' and `suppressable,' which we call \emph{capabilities}.
    \item A policy $P$ describing the set of all logs that are deemed legal.
\end{enumerate}
In the past, various logics have been used to describe policies to be enforced at runtime. In this paper, we will focus on Metric First-Order Temporal Logic (MFOTL)~\cite{Chomicki1995,Basin2015}, for which state-of-the-art enforcement tools exists~\cite{Hublet2022,Hublet2024}.

\newcommand{\kw}[1]{{\color{blue}\mathtt{#1}}}
\renewcommand{\neg}{\mathop{\kw{NOT}}}
\renewcommand{\land}{\mathop{\kw{AND}}}
\renewcommand{\lor}{\mathop{\kw{OR}}}
\newcommand{\limp}{\mathop{\kw{IMPLIES}}}
\renewcommand{\exists}{\mathop{\kw{EXISTS}}}
\renewcommand{\forall}{\mathop{\kw{FORALL}}}
\newcommand{\Next}{\mathop{\kw{NEXT}}}
\newcommand{\Prev}{\mathop{\kw{PREVIOUS}}}
\newcommand{\Once}{\mathop{\kw{ONCE}}}
\newcommand{\Historically}{\mathop{\kw{HISTORICALLY}}}
\newcommand{\Eventually}{\mathop{\kw{EVENTUALLY}}}
\newcommand{\Always}{\mathop{\kw{ALWAYS}}}

\begin{figure}[t]
\centering
\begin{tabular}{ll}
$e(t_1,\dots,t_k)$ & Event $e$ occurs in the log with arguments $t_1$, \dots, $t_k$\\
$\neg \varphi$ & $\varphi$ is not fulfilled\\
$\varphi \land \psi$ & Both $\varphi$ and $\psi$ are fulfilled\\
$\varphi \lor \psi$ & Either $\varphi$ or $\psi$ is fulfilled\\
$\varphi \limp \psi$ & If $\varphi$ is fulfilled, then $\psi$ is fulfilled\\
$\exists x.~\varphi$ & There exists some value of $x$ such that $\varphi$ is fulfilled\\
$\forall x.~\varphi$ & For any value of $x$, the formula $\varphi$ is fulfilled\\
$\Once \varphi$ & The formula $\varphi$ is fulfilled at some time-point in the past or present \\
$\Historically \varphi$ & The formula $\varphi$ is fulfilled at all time-points in the past or present \\
$\Eventually \varphi$ & The formula $\varphi$ is fulfilled at some time-point in the present or future \\
$\Always \varphi$ & The formula $\varphi$ is fulfilled at all time-points in the present or future
\end{tabular}
    \caption{Syntax and semantics of selected MFOTL operators.\label{fig:MFOTL}}
    \vspace{-3ex}
\end{figure}

A fragment of the syntax of MFOTL and its  intuitive semantics are described in Figure~\ref{fig:MFOTL}. 
For a formal account, we refer the reader to previous work~\cite{Basin2015,Hublet2024}. 
As an example, the policy $P_1$ can be expressed as MFOTL formula $\varphi_1$
\begin{align*}
    & \Always~(\\
    &\quad\forall app, data, user, purpose. \\
    &\quad\quad\mathsf{uses}(app,data,user,purpose) \limp\; \Once (\mathsf{consent}(user,app,purpose)))
\end{align*}
which reads ``at all times, for any $app$, $data$, $user$, and $purpose$, 
whenever the application $app$ uses the data $data$ of user $user$ for purpose $purpose$, then, at an earlier time-point
in the log, $user$ has given consent to $app$ to use their data for $purpose$.''

\paragraph{Properties of policies.} Given an ontology and assumptions on the causability and suppressability
or events, we say that a policy is
\begin{description}
 \item[enforceable] if there exists an enforcer ensuring that SuS is policy-compliant at all times;
 \item[transparently enforceable] if there exists an enforcer ensuring that SuS is policy-compliant at all times \emph{and} it causes or suppresses events \emph{only} when not causing or not suppressing them would lead to the policy being violated.
\end{description}

Assuming that data usage is suppressable, the policy $P_1$ above is enforceable:
the enforcer can simply prevent data usage. 
By considering an enforcer that prevents data usage 
\emph{only} when it is not preceded by consent, we see that $P_1$
is also transparently enforceable, as usage is suppressed only when this is necessary for compliance.
Our tool WhyEnf~\cite{Hublet2024} can decide whether an MFOTL formula 
is from some (transparently) enforceable fragment, and it is able to enforce all formulae it can identify as enforceable.

\section{Requirements and methodology}
\label{sec:methodology}

We describe list four key requirements for enforceable specifications of legal provisions and present our
iterative methodology to develop such a specification.

\paragraph{Requirements.} What properties should a specification of legal provisions amenable to RE
have? The first requirement is straightforward; it comes in two variants:
\begin{enumerate}[leftmargin=1cm]
    \item[R1.1)] The policy must be enforceable.
    \item[R1.2)] The policy must be transparently enforceable.
\end{enumerate}
Some policies are known to be enforceable, but not transparently enforceable~\cite{Hublet2024}. Hence, R1.2 might not always be 
achievable when R1.1 is.

Whether a policy is enforceable depends on its structure and on the causability and suppressability of the events it uses.
In order for a specification to be usable for RE, we want such assumptions to be realistic.
That is, we want to be able to effectively \emph{instrument} the SuS in such a way that the enforcer can cause 
or suppress the actions that causable and suppressable events respectively represent.  Moreover, we want SuS to accurately report all observable events to the enforcer. We call this property
\emph{instrumentability}. In practice, instrumentability can be assessed either generically for a family of systems
fulfilling certain (formal or informal) requirements, or specifically for a single SuS. The latter can be most
convincingly demonstrated by actually implementing the code that reports relevant events and executes the enforcer's commands. We obtain
\begin{enumerate}[leftmargin=0.75cm]
    \item[R2)] The system must be instrumentable based on the stated capabilities, i.e.:
    \begin{enumerate}
        \item[R2a)] The system must  report all actions that give rise to observable events.
        \item[R2b)] The system must support causation of all actions that give rise to causable events.
        \item[R2c)] The system must support suppression of all actions that give rise to suppressable events.
    \end{enumerate}
\end{enumerate}

Another key requirement for a specification of legal provisions is that is should capture an
acceptable interpretation of the law.
This is a qualitative requirement that can only be fully assessed through an interdisciplinary effort with legal experts. 
How literal and strict the interpretation should be---e.g., whether the policy should strictly follow the logical 
structure of 
the law or simply state some stronger condition that guarantees legality---will depend on the specific application.
In the following, we refer to this as the \emph{faithfulness} requirement. To obtain a faithful specification, not 
only does the policy need to be carefully formulated, but the ontology must 
also be designed in such a way that the semantics of each event aligns with legal definitions and are sufficiently precise to avoid an incorrect instrumentation. We thus get
\begin{enumerate}
    \item[R3)] The ontology and policy must faithfully capture the underlying legal provisions, i.e., all of the following conditions must hold:
    \begin{enumerate}
    \item[R3a)] The policy must capture an acceptable interpretation of the law,
    \item[R3b)] Every event in the ontology must come together with a clear documentation of its semantics,
    \item[R3c)] This semantics must be sufficiently precise so that the instrumentation of each action (logging and possibly causation/suppression of events) will be consistent with legal definitions and the expectations of legal experts.
    \end{enumerate}
\end{enumerate}
For consistency with previous work~\cite{DeYoung2010,Robaldo2020}, we will focus on a `literal' 
approach that closely follows the logical structure of existing legislation. Hence, R3a will be understood as
requiring a close logical correspondence with the law.

As formal specifications of legal provisions should serve as bridges between the legal and technical
communities, it is reasonable to require that these specifications be understandable by both
legal and technical experts. Yet,
accessibility
to an audience without \emph{any} exposure to formal reasoning is likely to be infeasible,
even when using a user-friendly surface representation~\cite{Bartolini2018}.
Fortunately, legal
experts involved in the assessment of formal specifications of software systems are generally
more technically experienced than their peers. They may additionally be offered some form
of training that should, however, not be excessively long or difficult. Those trained experts
are those we will refer to as `legal experts' in the rest of this paper.

Ensuring the \emph{understandability} of formal specifications is not only important for 
legal experts. Even technical experts' understanding of a policy may be 
influenced when the complexity of the policy grows. As the the representation 
of the same policy in various formal frameworks can be very different,
the understandability of a specification depends on the formalism used. 
But even within a given formalism, certain design decisions
such as, e.g., ensuring that the policy is written in a reasonably
concise way, can significantly improve understandability.

\begin{enumerate}[leftmargin=0.75cm]
\item[R4)] The policy must be understandable by both technical and legal experts, i.e., all of the following conditions must hold:
  \begin{enumerate}
  \item[R4a)] The logical formalism must be accessible to both technical and legal audiences assuming a moderate amount of training,
  \item[R4b)] The policy must be expressed in a clear and concise way,
  \item[R4c)] The complexity (size, logical structure) of the policy must not grow beyond what experts can comprehend.
  \end{enumerate}
\end{enumerate}

Only R1 can be automatically checked using a tool such as WhyEnf~\cite{Hublet2024}, 
while R2 can be demonstrated by either concrete implementation or high-level reasoning and R3--4 are qualitative. 
Figure~\ref{fig:requirements} summarizes the parts of the specification that each requirement concerns and 
which tool or expert(s) can assert its fulfillment.
\begin{figure}
\centering
\begin{tabular}{rcccl}
& \multicolumn{3}{c}{Involves...} & \\
\cline{2-4}
& Ontology & Capabilities & Policy & Assessment by...\\
\hline
R1 & $\checkmark$ & $\checkmark$ & $\checkmark$ & Tool (e.g.~\cite{Hublet2024}) \\
R2a & $\checkmark$ & $\checkmark$ & & Technical expert \\
R2b & $\checkmark$ & $\checkmark$ & & Technical expert \\
R2c & $\checkmark$ & $\checkmark$ & & Technical expert \\
R3a & $\checkmark$ & & $\checkmark$ & Legal expert \\
R3b & $\checkmark$ & & & Legal expert \\
R3c & $\checkmark$ & & & Legal and technical expert \\
R4a & & & $\checkmark$ & Legal and technical expert \\
R4b & & & $\checkmark$ & Legal and technical expert \\
R4c & & & $\checkmark$ & Legal and technical expert
\end{tabular}
\vspace{-1ex}
\caption{Summary of requirements\label{fig:requirements}}
\vspace{-3ex}
\end{figure}

\paragraph{Methodology.} We now present an iterative methodology to develop a specification fulfilling R1--4.
The flowchart of our methodology is shown in Figure~\ref{fig:methodology}.

Our methodology resembles pair programming~\cite{Merigoux2021} in which a technical expert (TE, e.g., a programmer or a logician) collaborates with a legal expert (LE) to develop an enforceable 
specification of a given set of legal provisions. The TE and LE start by selecting a formalism or a series of formalisms that fulfills
R4a. Using several formalisms that can be mechanically converted into each other can be meaningful when, e.g., the LE is 
more comfortable with a textual representation of the formula while the TE wishes a more mathematical representation~\cite{Bartolini2018}.
If previous work has already developed a partial specification of the law, then this existing specification can
be converted into a specification in the new formalism and serve as a starting point. 

After a possible conversion step, the TE and LE start by jointly drafting an ontology representing the law's main concepts and checking its compatibility 
with requirements R3b--c (clarity, precision). Next, they draft a policy representing the law's provisions and 
check its compatibility with requirements R3a and R4b--c (acceptable interpretation of the law, understandability). Then,
the TE sets the capacities of each event described in the ontology to fulfill R2a--c. Finally, they use a tool such as WhyEnf~\cite{Hublet2024} to
check whether the policy is enforceable (i.e., R1). If this is the case, then the TE and LE have successfully 
derived a specification that fulfills all of the stated requirements. Otherwise, the following can be attempted
to recover enforceability: (1) extend the capabilities (if possible) to observe, cause, or suppress more actions
(2) modify the policy (while preserving R3a and R4b--c) to make it enforceable (3) modify the ontology and/or the policy
(while preserving all requirements) to make the policy enforceable. We suggest to try these in the order of the
least required changes to the specification.

\tikzstyle{startstop} = [rectangle, rounded corners, minimum width=3cm, minimum height=1cm,text centered, draw=black, fill=red!30]
\tikzstyle{io} = [trapezium, trapezium left angle=70, trapezium right angle=110, minimum width=3cm, minimum height=1cm, text centered, draw=black, fill=blue!30]
\tikzstyle{process} = [rectangle, minimum width=3cm, minimum height=1cm, text centered, draw=black, fill=orange!30, yshift=-0.5cm]
\tikzstyle{decision} = [diamond, minimum width=3cm, minimum height=1cm, text centered, draw=black, fill=green!30]
\tikzstyle{arrow} = [thick,->,>=stealth]

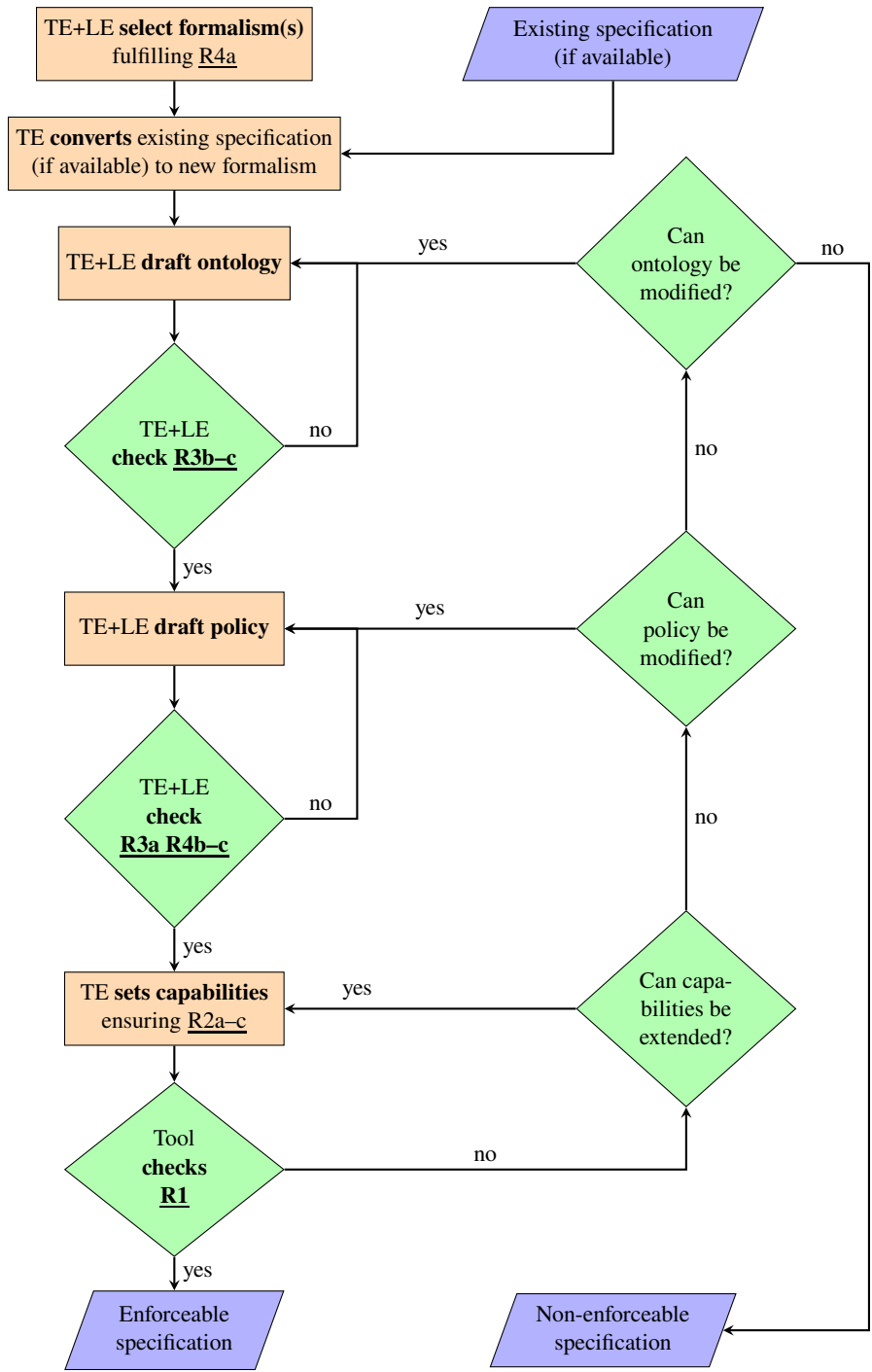
\begin{figure}
\centering
\begin{tikzpicture}
\node (formalism) [process, align=center] {TE+LE \textbf{select formalism(s)}\\fulfilling \underline{R4a}};
\node (ex) [io, right of=formalism, xshift=5cm, align=center] {Existing specification\\(if available)};
\node (conversion) [process, below of=formalism, align=center] {TE \textbf{converts} existing specification\\(if available) to new formalism};
\node (draftOntology) [process, below of=conversion] {TE+LE \textbf{draft ontology}};
\node (checkOntology) [decision, below of=draftOntology, yshift=-1.5cm, align=center] {TE+LE\\\textbf{check \underline{R3b--c}}};
\node (draftPolicy) [process, below of=checkOntology, yshift=-1cm, align=center] {TE+LE \textbf{draft policy}};
\node (checkPolicy) [decision, below of=draftPolicy, yshift=-1.6cm, align=center] {TE+LE \\ \textbf{check}\\\textbf{\underline{R3a R4b--c}}};
\node (draftCapabilities) [process, below of=checkPolicy, yshift=-1.1cm, align=center] {TE \textbf{sets capabilities}\\ ensuring \underline{R2a--c}};
\node (checkEnforceable) [decision, below of=draftCapabilities, yshift=-1.2cm, align=center] {Tool \\\textbf{checks}\\\textbf{\underline{R1}}};
\node (checkCapabilities) [decision, right of=draftCapabilities, align=center, xshift=6cm] {Can capa-\\bilities be\\extended?};
\node (checkPolicyUpdate) [decision, right of=draftPolicy, align=center, xshift=6cm] {Can\\policy be\\modified?};
\node (checkOntologyUpdate) [decision, right of=draftOntology, align=center, xshift=6cm] {Can\\ontology be\\modified?};
\node (output1) [io, below of=checkEnforceable, yshift=-1.2cm, align=center] {Enforceable \\ specification};
\node (output2) [io, right of=output1, xshift=5cm, align=center] {Non-enforceable \\ specification};

\draw[arrow] (formalism) to (conversion);
\draw[arrow] (ex) |- (conversion);
\draw[arrow] (conversion) to (draftOntology);
\draw[arrow] (draftOntology) to (checkOntology);
\draw[arrow] (checkOntology)  --  node[above,pos=0.5] {no} ++(2.5,0) |- (draftOntology);
\draw[arrow] (checkOntology) -- node[right,pos=0.5] {yes} (draftPolicy);
\draw[arrow] (draftPolicy) -- (checkPolicy);
\draw[arrow] (checkPolicy)  --  node[above,pos=0.5] {no} ++(2.5,0) |- (draftPolicy);
\draw[arrow] (checkPolicy) -- node[right,pos=0.5] {yes} (draftCapabilities);
\draw[arrow] (draftCapabilities) -- (checkEnforceable);
\draw[arrow] (checkEnforceable) -- node[right,pos=0.5] {yes} (output1);
\draw[arrow] (checkEnforceable) -| node[above,pos=0.25] {no} (checkCapabilities);
\draw[arrow] (checkCapabilities) -- node[above,pos=0.75] {yes} (draftCapabilities) ;
\draw[arrow] (checkCapabilities) --  node[right,pos=0.5] {no} (checkPolicyUpdate) ;
\draw[arrow] (checkPolicyUpdate) -- node[right,pos=0.5] {no} (checkOntologyUpdate);
\draw[arrow] (checkPolicyUpdate) -- node[above,pos=0.5] {yes} (draftPolicy);
\draw[arrow] (checkOntologyUpdate) -- node[above,pos=0.5] {yes} (draftOntology);
\draw[arrow] (checkOntologyUpdate) -- node[above,pos=0.5] {no} ++(2.5,0) |- (output2);
\end{tikzpicture}
\vspace{-1ex}
\caption{Flowchart of our methodology\label{fig:methodology}}
\end{figure}

\section{Case study}

\label{sec:case study}
In this section, we report on a preliminary case study in developing an enforceable GDPR specification using
the methodology in Section~\ref{sec:methodology}. This case study was conducted as part of one of the authors' Master's 
thesis~\cite{Kvamme2024}, using the existing specification from DAPRECO~\cite{Robaldo2020} as a starting point. 
While this preliminary case study did \emph{not} involve legal experts, we claim that it already demonstrates the potential
of our approach to improve over the state of the art. Namely, our case study allowed us to discover and correct
a number of inaccuracies in the existing specification, and generate a specification of single GDPR 
provisions that is directly amenable to formal reasoning using the WhyEnf enforcement tool~\cite{Hublet2024}.

\paragraph{Formalism selection and conversion.} 
We pick MFOTL as a formalism for specifying our policies as previous work~\cite{Arfelt2019,Hublet2023} has shown that MFOTL
provides the necessary expressivity for encoding (parts of) existing laws, and enforcement tools for MFOTL
are available~\cite{Hublet2022,Hublet2024}.

We first developed a series of algorithms to convert the Reified I/O Logic specification by Robaldo et al.~\cite{Robaldo2020} 
into an equivalent\footnote{In the course of developing this conversion algorithm, it became apparent that some of the logical formalisms used by Robaldo et al., especially reification, lacked formal semantics. A semantics thus had to be reconstructed based on textual descriptions from previous work.} MFOTL specification~\cite{Kvamme2024}, and applied it
to the DAPRECO knowledge base to obtain an MFOTL formula for each of 966 Reified I/O Logic formulae of DAPRECO.
Conversion involved transforming I/O rules into classical implications to be enforced; identifying the temporal patterns
encoded through reification; and rewriting these patterns using 
MFOTL operators. We also implemented a number of static checks
aimed at identifying incorrect syntactic patterns, such as unused variables.
The DAPRECO ontology, which to the best of our knowledge remained documented, was then extracted from the resulting MFOTL 
formulae.

\paragraph{Deriving an enforceable formal specification of Art. 7(1).} In the following, we describe in detail how
we derived an enforceable specification for Article 7(1) GDPR and corrected inaccuracies in an initial specification
derived from DAPRECO. We chose this article as it served as an
example to demonstrate DAPRECO's validation methodology~\cite{Bartolini2018,Bartolini2019}, which involved legal experts.
Article 7(1) GDPR states
\begin{quote}
    Where processing is based on consent, the controller shall be able to demonstrate that the data subject has consented to processing of his or her personal data.
\end{quote}
After an automated conversion to MFOTL, the DAPRECO version of this provision reads
\begin{align*}
\varphi_{7(1)}^D &= \Always~(\\
&\quad\! \exists ep,eau,edp,w,z,x,epu,c.~(\\
&\quad\quad\mathsf{PersonalDataProcessing}(ep, x, z) \land \mathsf{isBasedOn}(ep,epu) \\
&\quad\quad\land \mathsf{GiveConsent}(ehc,w,c) \land \mathsf{AuthorizedBy}(eau,epu,c) \\
&\quad\quad\land \mathsf{nominates}(edp,y,x) \land \mathsf{PersonalData}(z,w)\land \mathsf{Purpose}(epu))\\
&\quad\!\limp~(\exists ea, ed.~(\mathsf{AbleTo}(ea,y,ed) \land \mathsf{Demonstrate}(ed,y,ehc)))).
\end{align*}
English equivalent: ``{\itshape
Whenever personal data of an individual is processed by a processor nominated by a controller,
and the processing is based on a purpose, and the individual's consent authorizes the purpose (\emph{sic}!),
then the controller must be able to demonstrate that that consent has been given.
}''

\vspace{10pt}
The corresponding (undocumented) ontology contains all the events appearing in the formula. First, we need to 
reconstruct the precise semantics of the events it contains (R3b--c). For space reasons, we do not spell
out the semantics of all of these events here, but focus on $\mathsf{GiveConsent}$, $\mathsf{inBasedOn}$,
and $\mathsf{AuthorizedBy}$.
From its use in the specification, the semantics of
$\mathsf{GiveConsent}(ehc,w,c)$ can be reconstructed as ``the individual $w$ gives consent $c$; the individual's action of giving consent is given the unique identifier $ehc$.'' The meaning of
$\mathsf{isBasedOn}(ep,epu)$ and $\mathsf{AuthorizedBy}(eau,epu,c)$
is much less clear: the use of the variable $epu$ suggest the semantics
``the processing task $ep$ is \emph{based on the purpose} $epu$'' and
``[usage for] purpose $epu$ is authorized by consent $c$; this authorization relation is given the unique identifier $eau$.'' However, the law only refers to
``\emph{based on consent},'' not to ``based on a purpose'' and ``authorized by consent,''
and the consent action appears to be missing the identity of the controller.
While Robaldo et al. may have chosen this approach for consistency with the
earlier PrOnto ontology~\cite{Palmirani2018}, which has $\mathsf{isBasedOn}$, we see this part of the
ontology as diverging from the GDPR's concepts. We suggest to instead use two predicates
$\mathsf{isBasedOn}(ep,ehc)$ and $\mathsf{GiveConsent}(ehc,w,x,epu)$ with the semantics
``data processing $ep$ is based on consent action $ehc$'' and ''by consent action $ehc$, user $w$ grands
consent to $x$ to use her data for purpose $epu$,'' and to add an event $\mathsf{hasPurpose}(ep,epu)$
meaning ``the processing $ep$ has the purpose $epu$.'' The fact that $epu$ is always a purpose makes an\pagebreak[2]
additional $\mathsf{Purpose}(epu)$ event redundant in this case.
After modifying the ontology, our formula is
\begin{align*}
\varphi_{7(1)}^2 &= \Always~(\\
&\quad\! \exists ep,eau,edp,w,z,x,epu,c.~(\\
&\quad\quad\mathsf{PersonalDataProcessing}(ep, x, z) \land \mathsf{isBasedOn}(ep,ehc) \\
&\quad\quad\land \mathsf{GiveConsent}(ehc,w,x,epu) \land \mathsf{hasPurpose}(ep,epu)  \\
&\quad\quad\land \mathsf{nominates}(edp,y,x) \land \mathsf{PersonalData}(z,w))\\
&\quad\!\limp~(\exists ea, ed.~(\mathsf{AbleTo}(ea,y,ed) \land \mathsf{Demonstrate}(ed,y,ehc)))).
\end{align*}
English equivalent: ``{\itshape
Whenever personal data of an individual is processed by a processor nominated by a controller, and the 
processing is based on consent given by the user for the purpose of the current processing, then
the controller must be able to demonstrate that that consent has been given.
}''
\vspace{10pt}

Now, we perform the checks related to the policy. Does $\varphi^D_{7(1)}$ accurately reflect the letter of the 
law (R3a)? We claim that the temporal structure of the formula is not correct, as the specification 
states that an obligation to demonstrate consent only exists when the data processing 
($\mathsf{PersonalDataProcessing}$) and consent ($\mathsf{GiveConsent}$) are \emph{simultaneous}. 
This obviously goes against standard legal interpretations (see, e.g., \cite[Art. 7, Rn. 6]{Paal2021}).
Hence, we must correct $\varphi_{7(1)}^{2}$ by replacing $\mathsf{GiveConsent}(ehc,w,x,epu)$ by 
$\Once~\mathsf{GiveConsent}(ehc,w,x,epu)$. We call the resulting formula $\varphi_{7(1)}^3$
Note that this problem is independent of the mismatch between
GDPR concepts and the ontology that we corrected in the previous step. Is the formula clear (R4b--c)?
From a mathematical point of view, one potential source of confusion is that two variables ($ehc,y$) 
are implicitly universally quantified~\cite{Robaldo2020}. 
After adding explicit quantifiers (i.e., $\Always\forall ehc,y\dots$), we obtain
\begin{align*}
\varphi_{7(1)}^4 &= \Always\forall ehc,y.~(\\
&\quad\! \exists ep,eau,edp,w,z,x,epu,c.~(\\
&\quad\quad\mathsf{PersonalDataProcessing}(ep, x, z) \land \mathsf{isBasedOn}(ep,ehc) \\
&\quad\quad\land\,(\Once \mathsf{GiveConsent}(ehc,w,x,epu)) \land \mathsf{hasPurpose}(ep,epu)  \\
&\quad\quad\land \mathsf{nominates}(edp,y,x) \land \mathsf{PersonalData}(z,w))\\
&\quad\!\limp~(\exists ea, ed.~(\mathsf{AbleTo}(ea,y,ed) \land \mathsf{Demonstrate}(ed,y,ehc)))).
\end{align*}
English equivalent: ``{\itshape
Whenever personal data of an individual is processed by a processor nominated by a controller, and the 
processing is based on consent previously given by the user for the purpose of the current processing, then
the controller must be able to demonstrate that that consent has been given.
}''
\vspace{10pt}

We claim that $\varphi_{7(1)}^4$ is mathematically clear and concise, and that its complexity (a few lines) 
is reasonable. Making this formula understandable to non-technical experts does, however, require some
accurate higher-level representation, e.g., of the form proposed by Robaldo et al.~\cite{Robaldo2020}.

Next, we set the capabilities, assuming, as suggested in Section~\ref{sec:enforcement}, that 
data processing in our SuS can be subjected to prior approval by the enforcer. In this case, 
$\mathsf{PersonalDataProcessing}$ can be made observable and suppressable. The other relevant actions
can all be made observable by ensuring that the SuS can report the legal basis and purposes of processing, received consent, processor-controller delegation relationships, and its capacity to demonstrate consent
to the enforcer. This satisfies R2a--c. Finally, we use WhyEnf~\cite{Hublet2024} to check the enforceability
of $\varphi^4_{7(1)}$ when $\mathsf{PersonalDataProcessing}$ is suppressable. 
The check succeeds, guaranteeing transparent enforceability (R1.2): to ensure compliance with $\varphi^4_{7(1)}$ 
at all times, it necessary and sufficient to prevent personal data processing when the system can not
demonstrate that it has previously obtained consent.

\paragraph{Discussion.} We performed a similar effort for all provisions 
of Articles 5, 6, 7, 11, 12, 13, and 17 covered by DAPRECO. We discovered many issues related to unclear semantics of predicates, inaccurate modeling of time, unused variables, and lack of clarity~\cite{Kvamme2024}.

The fact that even the specification that Robaldo et al. validated with legal experts~\cite{Robaldo2020}
proved to be incorrect after a more precise analysis shows the benefits of our methodology and its
anchoring in formal methods, especially RE.
The criteria of ``accuracy, completeness, correctness, consistency, and conciseness''~\cite{Robaldo2020}
against which Robaldo et al. asked their experts to evaluate the policy failed to prevent the inaccuracies
we identified,
likely because
(1) the semantics of the ontology was never documented and evaluated together with the policy, and
(2) the representation of time in Reified I/O logic was too complex to be properly used even
by the experts themselves. 

Our conversion to MFOTL made many modeling mistakes apparent. Hence, another take-away 
is that MFOTL is a promising specification language to formalize legal provisions, allowing
for expressive first-order modeling with a transparent encoding of time.

\section{Conclusion and open questions}

In this paper, we have presented four requirements and a new methodology for developing
enforceable formal specifications of privacy laws. We have then reported on a preliminary
case study focusing on selected provisions of the General Data Protection Regulation (GDPR).
In the course of our case study, we have identified several inaccuracies in an existing
GDPR specification. Overall, our case study demonstrates the benefits of our methodology and
suggests Metric First-Order Temporal Logic (MFOTL) as a promising  language
for formalizing laws.

In order to obtain an enforceable, state-of-the-art specification that can serve as
a reference for both computer scientists and legal experts, our preliminary
case study needs to be extended in two directions: by involving several
technical and legal experts, and by extending the scope of the formalization effort to a larger
fragment of the GDPR---ideally covering all provisions that regulate computer systems' behavior.
To allow for efficient collaboration between legal and technical experts in this context,
relying a more user-friendly specification language with some temporal features (rather than
just MFOTL) is indispensable. We plan to develop such a language as part of our future
work.

Another open question is how to \emph{refine} general specifications of
the `literal' kind we discussed into simpler, more concrete specifications tailored to
guarantee the compliance of specific systems. While a rich theory of refinement exists within formal methods, we are not aware of any previous work that would apply these
techniques in a legal context.

Last but not least, our preliminary results and the inaccuracies we identified in previous work call for the development of more systematic methodology for the joint assessment of legal compliance by both legal and technical experts. In general, technical experts alone cannot assert the compliance of a system with legal requirements. But neither can legal experts if their understanding of the system's behavior is not informed by trustworthy technical experts' knowledge. As a result, only a joint assessment of compliance is ever possible. This raises fundamental theoretical and practical questions than can be interesting to both the technical and legal communities.

\label{sec:conclusion}
%
%
%
\bibliographystyle{splncs04}
\bibliography{main}

\end{document}